\documentclass[12pt]{iopart}
\usepackage{iopams}
\usepackage{graphicx}
\usepackage[T1]{fontenc}
\usepackage{ae}
\usepackage[latin1]{inputenc}
\newcommand{\Eins}
           {\;\smash{\raisebox{-0.5ex}{$\!\!\stackrel{\!\mbox{1}
            \hspace{-0.4ex}\rule[0.0ex]{0.06ex}{1.60ex}}{ }$}}}

\newtheorem{lemma}{Lemma}
\newtheorem{defi}{Definition}
\newtheorem{prop}{Proposition}

\newtheorem{theorem}{Theorem}
\newtheorem{cor}{Corollary}

\begin{document}

\title[Dimerized ground states]
{Spin systems with dimerized ground states }

\author{Heinz-J\"urgen Schmidt\dag
\footnote[3]{Correspondence should be addressed to
hschmidt@uos.de} }

\address{\dag\ Universit\"at Osnabr\"uck, Fachbereich Physik,
Barbarastr. 7, 49069 Osnabr\"uck, Germany}

\begin{abstract}
In view of the numerous examples in the literature it is attempted
to outline a theory of Heisenberg spin systems possessing
dimerized ground states (``DGS systems") which comprises all known
examples. Whereas classical DGS systems can be completely
characterized, it was only possible to provide necessary or
sufficient conditions for the quantum case. First, for all DGS
systems the interaction between the dimers must be balanced in a
certain sense. Moreover, one can identify four special classes
of DGS systems: (i) Uniform pyramids, (ii) systems close to
isolated dimer systems, (iii) classical DGS systems, and (iv),
in the case of $s=1/2$, systems of two dimers satisfying four
inequalities. Geometrically, the set of all DGS systems may be
visualized as a convex cone in the linear space of all exchange
constants. Hence one can generate new examples of DGS systems by
positive linear combinations of examples from the above four
classes.
\end{abstract}

\pacs{75.10.b, 75.10.Jm}

\submitto{\JPA}

\maketitle

\section{Introduction}
\label{sec1}
 Spin systems with exact ground states are rare and
hence have found considerable interest. A trivial case is a system
of $N$ unconnected antiferromagnetic (AF) dimers which has the
product $\Phi$ of the individual dimer ground states as its unique
ground state. An interaction between the dimers would in general
perturb the ground state, but, interestingly, for certain
interactions $\Phi$ remains a ground state. In these cases the
perturbational corrections of all orders will vanish; the
interaction between the
dimers is, so to speak, frozen for low temperatures.\\

Examples of these systems which minimize their energy for a
product state $\Phi$ of dimer ground states (``DGS systems") have
been constructed and studied in dozens of papers. Sometimes DGS
systems are also referred to as ``valence bond" (VB) systems, or,
more generally, as  ``resonating valence bond" (RVB) systems if
superpositions of VB states are involved. Here I can only mention
a small selection of this literature. In the seminal papers of
Majumdar and Ghosh \cite{MajumdarG69a}\cite{MajumdarG69b} even $s=1/2$ rings
are considered with constant NN and NNN interactions of relative
strength $2:1$ which possess two different DGS according to the
shift symmetry of the ring. More precisely, in the second of these
papers \cite{MajumdarG69b} the authors proved as ``an interesting
by-product" that $\Phi$ is an eigenstate of the Hamiltonian and
conjectured it being a ground state due to numerical studies up to
10 spins. One year later, Majumdar \cite{Majumdar70} mentions a proof of
the DGS property
for Majumdar-Ghosh rings given in a private communication by J.~Pasupathy.\\

The generalization of these results to arbitrary spin quantum
numbers $s$ is due to Shastry and Sutherland \cite{ShastryS81a}. A
different generalization of Majumdar-Ghosh rings has been given by
Pimpinelli \cite{Pimpinelli91} and re-discovered by Kumar
\cite{Kumar02} who extended the coupling within the $s=1/2$ ring
to $2n$-nearest neighbors with strengths $J_1=2n, J_2=2n-1,\ldots,
J_{2n}=1$. Though it might not be adequate to call such a spin
system still a ``ring". Already the ring with NNN interaction
could better be viewed as a ``ladder". Ladders with DGS property
have also been studied in \cite{BoseG93} and \cite{BoseC02}. Other
one-dimensional models with DGS are certain
dimer-plaquette chains, see \cite{IvanovR97} and \cite{RichterIS98}.\\

Another family of two-dimensional DGS systems can be traced back
to the work of Shastry and Sutherland \cite{ShastryS81b} on square
lattices with alternating diagonal bonds for every second square
and arbitrary $s$. These authors also suggest to classify DGS
states as a ``spin liquids" due to their short range correlation.
The S(hastry)S(utherland) model is physically realized in $Sr Cu_2
(BO_3)_2$, see \cite{KageyamaYSMOKKSGU99} and
\cite{MiyaharaU99}.\\

Generalizations of the SS model are possible by introducing,
additional to the nearest neighbor (NN) and diagonal (D) bonds,
new types of interaction, namely next nearest neighbor (NNN),
knight's-move-distance-away (KM) and further-neighbor-diagonal
(FND), see \cite{BoseM91},\cite{BhaumikB95} and \cite{Bose92b}.
The dimerized state $\Phi$ is then an eigenstate of the
Hamiltonian if the five coupling constants involved satisfy
\begin{equation}\label{1.1}
 J_1:J_2:J_3:J_4:J_5 = 1:1:\frac{1}{2} :\frac{1}{2}:\frac{1}{4}
 \;.
\end{equation}
If an inhomogeneous NN coupling is chosen, namely $J_0$ along the
dimer bonds and $J_1$ for the remaining NN, the choice
\begin{equation}\label{1.2}
J_0:J_1 = 7:1
\end{equation}
yields a DGS for the generalized SS model, see \cite{BhaumikB95}.\\

Finally, I mention generalizations of the SS model to arbitrary
dimensions \cite{SurendranS02} and by superpositions of uniform
pyramids ($s=1/2$) constructed by Kumar \cite{Kumar02} which will
be reconsidered in section
\ref{sec3.1}. For further related examples, see also \cite{Bose02}.\\

In view of the abundance of examples of DGS systems in the
literature I am not primarily interested in new examples but
will try to characterize the class of {\it all} examples.
Unfortunately, I have achieved a complete characterization only
in the case of classical spin systems. This and the partial
results for quantum systems are contained in sections 2 and 3.
After the general definitions (subsection 2.1) a
necessary condition is formulated (theorem 1 in subsection 2.2).
It says that $\Phi$ will be an eigenstate of the spin Hamiltonian iff a certain
balance condition for the four coupling constants between any two
dimers is satisfied. For classical spin systems this balance
condition can be strengthened to the condition of uniform coupling
between any pair of dimers (theorem 2). \\

In section 3 I will give sufficient conditions for DGS
systems and enumerate four special classes of examples. As
mentioned above, systems of these classes can be superposed by
positive linear combinations in order to form new DGS systems.
Here we identify a spin system with its matrix ${\Bbb J}$ of
exchange parameters and thus understand ``addition" of systems as
the addition of the corresponding matrices. Hence classical spin
systems and quantum spin systems with any $s$ are not
distinguished at the level of ${\Bbb J}$-matrices, but, of course,
the definition of classical and quantum DGS systems is different.\\

Subsection 3.1 describes ``uniform pyramids" which are systems of
$N$ dimers with uniform coupling between all $2N$ spins (the base
of the pyramid) plus an extra dimer one spin of which (the vertex
of the pyramid) is uniformly coupled to the other $2N$ spins.
These pyramids are DGS systems for arbitrary $s$ if the uniform
coupling constants are suitably chosen. Another class of DGS
systems is provided by small neighborhoods of unconnected dimer
systems (subsection 3.2). The radius $r$ of the neighborhood
decreases with $s$. Although $r$ is not the optimal value, there
seems to be a trend that the class of DGS systems is shrinking if
$s$ increases. This phenomenon can be illustrated by examples, see
section 4, but has not yet been strictly proven.\\

For small $N$ and $s$ the class of DGS systems can, in principle,
be explicitly determined. The method is sketched in subsection
\ref{sec3.3} and the result for $N=2, s=1/2$ is given in the form
of four inequalities for polynomial functions of the involved four
coupling constants. Recall that for a classical DGS system the
coupling between any two pairs of dimers must be uniform. Hence it
is possible to encode the structure of such a system by an
$N\times N$ matrix ${\Bbb G}$ instead of the $2N\times 2N$ matrix
${\Bbb J}$. Then it can be shown that the system is a classical
DGS system iff this matrix ${\Bbb G}$ is positive semi-definite,
i.~e.~iff all its eigenvalues (or all its principal minors) are
non-negative (theorem 3 in subsection 3.4). Moreover, if a
coupling matrix ${\Bbb J}$ belongs to a classical DGS system, then
it also belongs to a quantum DGS system for all values of $s$.
Hence theorem 3 defines a forth class of special DGS systems.\\

In section 4 I will consider two examples. The first one
(subsection 4.1) consists of two dimers which are weakly coupled
in a balanced but not uniform way. If $\epsilon(s)$ is the maximal
coupling strength such that the quantum system with spin quantum
number $s$ is still DGS, then it follows that
$\epsilon(s)\rightarrow 0$ for $s \rightarrow \infty$ since the
classical system is not DGS for all $\epsilon>0$.\\

The second example (subsection 4.2) consists of three dimers and,
due to symmetry assumptions, normalization and the balance
condition, two independent coupling constants, say $J_2$ and
$J_4$. It is still possible to exactly calculate the convex set of
DGS systems in the $J_2-J_4-$plane and to illustrate the subsets
defined in section 3 for this example.\\

For the sake of readability of the paper all proofs of
the previously formulated theorems and propositions are deferred to section 5.
In section 6 we investigate the geometric structure of the set
${\cal C}_\Phi$ of DGS systems, represented as the set of the
corresponding ${\Bbb J}$-matrices. It is easily shown that
${\cal C}_\Phi$ is a {\it proper, convex, generating cone} in the linear
space of all symmetric matrices satisfying the balance condition.
Moreover, we will see that the interior points of ${\cal C}_\Phi$ are
exactly those systems for which $\Phi$ is the unique ground state.
Systems at the boundary of ${\cal C}_\Phi$ have degenerate
ground states. In particular, the {\it faces} of ${\cal C}_\Phi$
consist of all DGS systems having the same eigenspace of ground
states. We close with a summary (section 7).

\section{Definitions and necessary conditions for DGS systems}
\label{sec2}
\subsection{Definitions}
\label{sec2.1} We consider systems of $2N$ spins with
one and the same individual spin
quantum number $s$ which are grouped into $N$ fixed pairs
(``dimers"). To indicate this grouping the spins will be denoted
by indices $\mu =(i,\delta)$ where $i=1,\ldots, N$ is the dimer
index and $\delta = 0,1$ distinguishes between the two spins
belonging to the same dimer. Further we consider Heisenberg
Hamiltonians
\begin{equation} \label{2.1.1}
H({\Bbb J})= \sum_{\mu\nu}J_{\mu\nu}\bi{s}_\mu\cdot \bi{s}_\nu
\;,
\end{equation}
where $\bi{s}_\mu=(s_\mu^{(1)},s_\mu^{(2)},s_\mu^{(3)})$ denotes
the $\mu$-th spin observable and ${\Bbb J}$ the $2N\times
2N$-matrix of real exchange parameters or coupling constants
$J_{\mu\nu}$.
All operators act on a $(2s+1)^N$-dimensional Hilbert space
${\cal H}=\bigotimes_{\mu=1}^{2N}{\cal H}_\mu$.
If the spin quantum number $s$ is fixed, we may
identify a spin system with its matrix ${\Bbb J}$.\\

Note that the exchange parameters $J_{\mu\nu}$ are not uniquely
determined by the Hamiltonian $H({\Bbb J})$ via (\ref{2.1.1}).
Different choices of the $J_{\mu\nu}$ leading to the same $H({\Bbb
J})$ will be referred to as different "gauges". We will adopt the
following gauges: First, the antisymmetric part of ${\Bbb J}$ does
not occur in the Hamiltonian (\ref{2.1.1}). Hence we will follow
common practice and choose $J_{\mu\nu}=J_{\nu\mu}$. Thus $\Bbb J$
is a real symmetric matrix . Second, since  ${\bf s}_\mu\cdot{\bf
s}_\mu =s(s+1)\;{\Eins}\;$ we may choose arbitrary diagonal
elements $J_{\mu\mu}$ without changing $H({\Bbb J})$, as long as
their sum vanishes, $\mbox{Tr}\;{\Bbb J}=0$. The usual gauge
chosen throughout the literature is $J_{\mu\mu}=0,\;
\mu=1,\ldots,N,$ which will be called the ``zero gauge". In this
article, however, we will choose the
``homogeneous gauge", which is defined by the condition that the
row sums will be independent of $\mu$, see also \cite{SchmidtS02}:
\begin{equation}\label{2.1.2}
j\equiv J_\mu \equiv \sum_\nu  J_{\mu\nu}
\end{equation}
Note that
the eigenvalues of ${\Bbb
J}$ may non--trivially depend on the gauge.\\

For any dimer with index $i$ let $[i0,i1]$ denote the ground state
of the AF dimer $\bi{s}_{i0}\cdot\bi{s}_{i1}$ which is unique up
to a phase and can be written in the form
\begin{equation}\label{2.1.3}
[i0,i1]= \frac{1}{\sqrt{2s+1}}\sum_{m=-s}^s (-1)^{s-m}|m,-m\rangle
\;,
\end{equation}
using the eigenbasis $|m\rangle,\; m=-s,\ldots,s$ of
$\bi{s}_\mu^{(3)}$. The ground state of a system of $N$
unconnected AF dimers is the product state
\begin{equation}\label{2.1.4}
\Phi^s \equiv \bigotimes_{i=1}^N [i0,i1] \;,
\end{equation}
called the {\it dimerized state}; it has the total spin quantum
number $S=0$. A system ${\Bbb J}$ is said to admit dimerized
ground states (DGS), or to have the DGS property, iff $\Phi^s$ is
a ground state of $H({\Bbb J})$, i.~e.~iff
\begin{equation}\label{2.1.5}
\langle \Phi^s | H({\Bbb J}) \Phi^s\rangle
\le
\langle \Psi |H({\Bbb J}) \Psi\rangle
\end{equation}
for all $\Psi\in{\cal H}$ with $||\Psi||=1$.  Let ${\cal
C}_\Phi^s$ denote the set of all spin systems ${\Bbb J}$ with the
DGS property. If the quantum number $s$ is understood,
we suppress it and write simply $\Phi$ and ${\cal C}_\Phi$. \\

Analogous definitions hold for the classical case:
Here the spin observables $\bi{s}_\mu^{\small{cl}}$ are
unit vectors, $H({\Bbb J})^{\small{cl}}$ is the Hamiltonian function,
defined on the $2N$-fold Cartesian product of unit spheres
\begin{equation}\label{2.1.6}
{\cal P} \equiv {\begin{array}{c}
{\scriptstyle N}\\
{\mbox{\Large\sf X}}
\\^{\scriptstyle \mu=1}\end{array}}
{\cal S}_{(\mu)}^2 \;,
\end{equation}
and $\Phi^{\small{cl}}\subset {\cal P}$ is the set of all
spin configurations satisfying
\begin{equation}\label{2.1.7}
\bi{s}_{i0} + \bi{s}_{i1} = \bi{0}
\;\mbox{ for all } i=1,\ldots,N
\;.
\end{equation}
Note that  $\Phi^{\small{cl}}$ as well as $\Phi^{\small{s}}$
are invariant under rotations. ${\Bbb J}$ is said to have the
classical DGS property iff the minimum of $H({\Bbb J})$ is assumed for all
$\bi{s}\in \Phi^{\small{cl}}$. In this case we write
${\Bbb J}\in {\cal C}_\Phi^{\small{cl}}={\cal C}_\Phi^{\infty}$.

\subsection{Necessary conditions for DGS systems}
\label{sec2.2}
Whereas a complete characterization of  ${\cal C}_\Phi^s$ seems to be
possible only for small $N$ and $s$ or for the classical case $s=\infty$,
one can prove a number of partial results, either necessary or sufficient
conditions for ${\Bbb J}\in {\cal C}_\Phi^s$. \\

\begin{figure}
  \includegraphics[width=13cm]{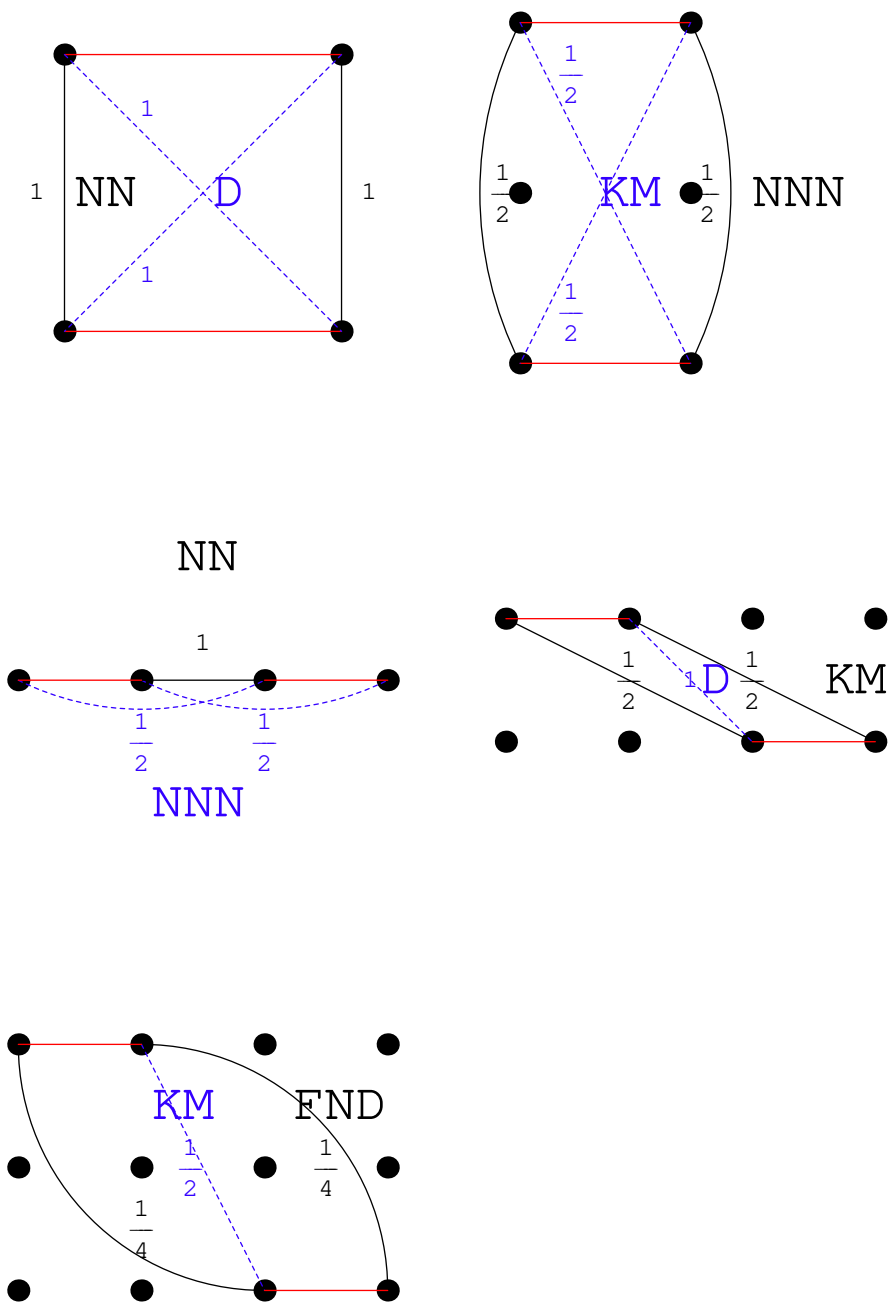}
\caption{\label{fig01}Decomposition of the generalized SS model \cite{BoseM91}
\cite{BhaumikB95} \cite{Bose92b} into pairs of dimers satisfying
(\ref{2.2.1}). The types of interaction are explained in the Introduction
}
\end{figure}

Necessary conditions of a rather trivial kind can be obtained
whenever one finds a state $\Psi\in{\cal H}$ such that the r.~h.~s.~of
(\ref{2.1.5}) can be explicitely calculated. Less trivial is the following
\begin{theorem}
\label{T1}
$\Phi$ is an eigenstate of $H({\Bbb J})$ iff
\begin{equation}\label{2.2.1}
J_{i0,j0}+ J_{i1,j1}=J_{i0,j1}+J_{i1,j0}
\end{equation}
for all $i<j=2,\ldots,N$.
\end{theorem}
This theorem says that for a DGS system the interaction between
any pair of dimers has to be balanced in a certain sense: The strength
of the inter-dimer parallel bonds must equal the strength of the diagonal bonds,
but the strength of the dimer bonds may be arbitrary.
For example, a spin square is never DGS because there are no diagonal
bonds at all.\\
To give another example, is easy to derive from theorem \ref{T1}
the condition (\ref{1.1}) for $\Phi$ being an eigenstate of the
generalized SS model by superposing five suitable
dimer pairs of the square lattice, see figure \ref{fig01}.\\

If $\Phi$ is an eigenstate of $H({\Bbb J})$, it is straight
forward to calculate the corresponding eigenvalue, since
$\langle\Phi|\bi{s}_{i\delta}\cdot\bi{s}_{j\epsilon}|\Phi\rangle=0$
for $i\neq j$, see section \label{sec5.1}.
\begin{cor}
\label{C1}
If $\Phi$ is an eigenstate of $H({\Bbb J})$, then
\begin{equation}\label{2.2.2}
H({\Bbb J})\Phi = -2s(s+1)\sum_{i=1}^N J_{i0,i1}\; \Phi
\;.
\end{equation}
\end{cor}
Because (\ref{2.2.1}) is a linear equation, the set of all real,
symmetric, traceless $2N\times 2N$-matrices ${\Bbb J}$ satisfying
(\ref{2.1.2}) and (\ref{2.2.1}) will be a linear space of
dimension $\frac{1}{2}N(3N-1)$, denoted by ${\cal J}_\Phi$. Further
note that the ground state condition (\ref{2.1.5}) is conserved
under positive linear combinations of ${\Bbb J}$'s. Hence the set
${\cal C}_\Phi$ of DGS systems will form a {\it convex cone}
embedded in the linear space ${\cal J}_\Phi$. Further geometrical
properties of ${\cal C}_\Phi$ will be discussed in section
\ref{sec6}.\\

Another necessary condition for the DGS property is the
following:
\begin{prop}
\label{P1}
If ${\Bbb J}\in {\cal C}_\Phi$ then $J_{i0,i1}\ge 0$ for all $i=1,\ldots,N$.
\end{prop}
Note, however, that the coupling {\it between} the dimers can be
negative and nevertheless the system may have a DGS. For example, this
may happen for systems close to unconnected dimer systems, see proposition
\ref{P3}.\\

In the classical case we have similar but stronger results: The
balance condition (\ref{2.2.1}) can be strengthened to a uniform
coupling condition:
\begin{theorem}
\label{T2}
If ${\Bbb J}\in {\cal C}_\Phi^{\small{cl}}$  then
\begin{equation}\label{2.2.3}
J_{i0,j0}= J_{i0,j1}=J_{i1,j0}=J_{i1,j1}\equiv\epsilon_{ij}
\end{equation}
and
\begin{equation}\label{2.2.4}
J_{i0,i1}\ge 0
\end{equation}
for all $i<j=2,\ldots,N$.
\end{theorem}
Consequently, we will denote by  ${\cal J}_\Phi^\infty\equiv{\cal
J}_\Phi^{\small{cl}}$ the linear space of all real, symmetric,
traceless $2N\times 2N$-matrices ${\Bbb J}$ satisfying
(\ref{2.1.2}) and (\ref{2.2.3}).

\section{Sufficient conditions for DGS systems}
\label{sec3} We consider four classes of DGS systems which will
hopefully cover all examples known from the literature by means of
positive linear combinations of ${\Bbb J}$-matrices. Note that
also DGS systems with a different number of dimers can be
superposed in this sense.
\subsection{Uniform pyramids}
\label{sec3.1}

\begin{figure}
  \includegraphics[width=10cm]{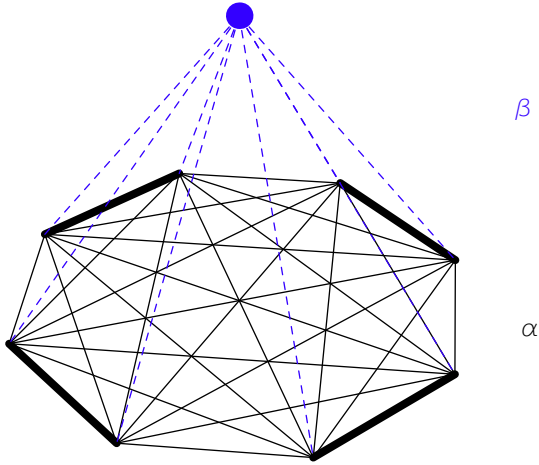}
\caption{\label{fig02}The uniform pyramid is a DGS system if $\alpha,\beta$
satisfy (\ref{3.1.2}).}
\end{figure}

A {\it uniform pyramid} consists of $N$ dimers with uniform
coupling between all $2N$ spins plus another uniform coupling with
one extra spin, see figure \ref{fig02}. This extra spin with
indices $(N+1,0)$ is considered as a part of another $N+1$-th
dimer, in order to make it possible to define the dimerized state
$\Phi$. More precisely, we require
\begin{eqnarray}\label{3.1.1a}
J_{i\delta,j \epsilon} &=& \alpha>0 \mbox{ for all }i,j=1,\ldots,N
\mbox{ and } \delta,\epsilon\in \{0,1\}\\   \nonumber &&\mbox{
except for }(i,\delta)=(j,\epsilon)\;,\\ \label{3.1.1b}
J_{i\delta,N+1\; 0} &=& \beta>0 \mbox{ for all }i=1,\ldots,N
\mbox{ and }\delta\in \{0,1\}
\end{eqnarray}
Then the following holds:
\begin{prop}
\label{P2}
Let ${\Bbb J}$ be a uniform pyramid such that
\begin{equation}\label{3.1.2}
\frac{\beta}{\alpha} \le
\left\{
\begin{array}{cll}
 \frac{1}{s+1} & \mbox{if}&  s>\frac{1}{2} \\
 1 & \mbox{if}&  s=\frac{1}{2}\;.
\end{array}
\right.
\end{equation}
Then ${\Bbb J}\in {\cal C}_\Phi$\;.
\end{prop}
DGS systems generated by uniform pyramids for $s=1/2$ have been
considered by Kumar \cite{K}. The $s=1/2$ Majumdar-Ghosh ring
\cite{MajumdarG69b} can be viewed as a superposition of $N$
uniform triangles, similarly the
SS model \cite{ShastryS81b}.\\

A superposition of uniform pyramids always yields DGS systems with $J_{\mu\nu}\ge 0$.
But this condition is not necessary, as can be seen by the next class of DGS systems.

\subsection{Neighborhoods of unconnected dimer systems}
\label{sec3.2} For matrices ${\Bbb J}\in{\cal C}_\Phi$ we define
its {\it spin modulus}
\begin{equation}\label{3.2.1}
{\cal x}({\Bbb J}) \equiv |j_{\small{min}}|
\end{equation}
as the absolute value of the lowest eigenvalue
$j_{\small{min}}$ of ${\Bbb J}$.
It has similar properties as a matrix norm. For
example, $j_{\small{min}}=0$ implies that all eigenvalues of
${\Bbb J}$ vanish, since $\mbox{Tr }{\Bbb J} =0$, and hence
${\Bbb J}=0$. But since, in general, ${\cal x}(-{\Bbb J})\neq {\cal x}({\Bbb J})$,
the spin modulus will not be a norm. Nevertheless it can  be used to
define neighborhoods of matrices ${\Bbb J}\in{\cal C}_\Phi$
because it can easily be shown that
\begin{equation}\label{3.2.1a}
\frac{1}{2N-1}\|{\Bbb J}\| \le {\cal x}({\Bbb J}) \le \|{\Bbb J}\|
\end{equation}
holds, where $\|{\Bbb J}\|\equiv \max \{\;\|{\Bbb J} x\| \;|\;\|x\|=1\;\}$
denotes the so-called {\it spectral norm}, see \cite{Lancaster69}.
\\

Let $\stackrel{\circ}{\Bbb J}$ denote the matrix of an unconnected AF dimer
system, i.~e.~$\stackrel{\circ}{J}_{i0,i1}>0$ for all $i=1,\ldots,N$ and all other non-diagonal
matrix elements vanish. Of course,  $\stackrel{\circ}{\Bbb J}\in{\cal C}_\Phi$.
But also a small neighborhood of $\stackrel{\circ}{\Bbb J}$ still
consists of DGS systems. More precisely:
\begin{prop}
\label{P3} Let  $\stackrel{\circ}{\Bbb J}$ be an unconnected dimer
system and $\lambda=\min\{ J_{i0,i1}|i=1\ldots N\}$. Further let
${\Bbb J}=\stackrel{\circ}{\Bbb J}+\Delta,\; \Delta\in{\cal
J}_\Phi$ such that
\begin{equation}\label{3.2.2}
{\cal x}(\Delta) \le \frac{\lambda}{N s(s+1)} \;.
\end{equation}
Then ${\Bbb J}\in{\cal C}_\Phi$.
\end{prop}
This proposition implies that the cone ${\cal C}_\Phi$ generates
${\cal J}_\Phi$, i.~e.~${\cal J}_\Phi={\cal C}_\Phi-{\cal
C}_\Phi$. Hence it is not possible to find a smaller subspace of
${\cal J}_\Phi$ which already contains all DGS systems.
This stands in contrast to the
classical case, see theorem \ref{T2}. Moreover, the $s$-dependence
of the bound in (\ref{3.2.2}) supports the conjecture that the
cones ${\cal C}_\Phi^s$ shrink with increasing $s$.

\subsection{The case $s=1/2$}
\label{sec3.3}
In principle, the cone ${\cal C}_\Phi$ could be exactly determined
as follows: Calculate the characteristic polynomial
$p(\lambda)=\det (H({\Bbb J})-\lambda\; \Eins)$  of
$H({\Bbb J}),\;{\Bbb J}\in{\cal J}_\Phi $.
Its roots are the real eigenvalues $E_\nu$ of $H({\Bbb J})$;
one of them, say $E_0$, will be the eigenvalue of $H({\Bbb J})$
w.~r.~t.~$\Phi$. Factor $p(\lambda)=q(\lambda)(\lambda-E_0)$. Then
${\Bbb J}\in {\cal C}_\Phi$ is equivalent to the condition that
$E_0\le$ any root of $q(\lambda)$. One can
easily find criteria for this inequality which do not assume that
the roots of $q(\lambda)$ are known. Consider, for example,
the simple case of $q$ being quadratic,
say $q(\lambda)=(\lambda-a)(\lambda-b),\; a<b$.
Then obviously $E_0 \le a$ iff $q(E_0)\ge 0$ and $q'(E_0)\le 0$.
More generally, one can prove the following:
\begin{lemma}
\label{L1}
Let $q(\lambda)=\sum_{\ell=0}^K a_\ell \lambda^\ell,\; a_K>0,$
have only real roots. Then $E_0\le$ any root of $q(\lambda)$
iff
\begin{equation}\label{3.3.1}
(-1)^{K+n}q^{(n)}(E_0) \ge 0 \mbox{ for all } n=0,\ldots,K-1\;.
\end{equation}
\end{lemma}
Thus it is possible to explicitely determine  ${\cal C}_\Phi$
by $K=(2s+1)^{2N}-1$ inequalities without calculating the $E_\nu$.
Unfortunately, it is practically impossible
to calculate $p(\lambda)$ for general ${\Bbb J}$,
even by using computer algebra software,
except for small values of $N$ and $s$. \\

I have determined ${\cal C}_\Phi$ by this method for the
simplest case of $N=2$ dimers and $s=1/2$. The result is the
following:
\begin{prop}
\label{P4}
Let $N=2$ and $s=1/2$ and ${\Bbb J}\in{\cal J}_\Phi $. Rewrite the dimer indices as\\
$(1,0)\equiv 0,\;(1,1)\equiv 1,\;(2,0)\equiv 2,\;(2,1)\equiv 3\;$.
\\
Then ${\Bbb J}\in {\cal C}_\Phi$ iff the following four inequalities hold:
\begin{eqnarray}\label{3.3.2a}
J_{01}+J_{23}-J_{02}-J_{13} &\ge& 0 \;,\\  \label{3.3.2b}
2(J_{01}+J_{23})+ J_{02}+J_{13} &\ge& 0 \;,\\  \nonumber
2J_{01}^2J_{23}-(J_{12}-J_{13})^2J_{23}- &&\\    \label{3.3.2c}
J_{01}((J_{02}-J_{12})^2+(J_{02}+J_{13})J_{23}-2J_{23}^2) &\ge& 0
\;,\\  \nonumber 2(J_{01}^2+J_{12}(J_{02}+J_{13})+J_{23}^2) &&\\
\label{3.3.2d} -J_{02}^2 -2
J_{12}^2-J_{13}^2-J_{01}(J_{02}+J_{13}+6 J_{23})
-(J_{02}+J_{13})J_{23} &\ge& 0 \;.
\end{eqnarray}
\end{prop}
We note that the inequalities in proposition \ref{P4}
can be written in a hierarchical order which makes it more
convenient to produce examples of DGS systems.
To this end we define
\begin{defi}
\label{D1}
\begin{eqnarray}\label{3.3.3a}
p_1 &\equiv& J_{01}+J_{23} \\ \label{3.3.3b}
p_2 &\equiv& J_{02}+J_{13} \\ \label{3.3.3c}
p_3 &\equiv& J_{12}-\frac{p_2}{2} \\ \label{3.3.3d}
p_4 &\equiv& J_{02}-\frac{p_2}{2} \\ \label{3.3.3e}
p_5 &\equiv& J_{01}
\end{eqnarray}
\end{defi}
Then the inequalities (\ref{3.3.2a})  to (\ref{3.3.2d}) are
equivalent to the following:

\begin{eqnarray}\label{3.4.3a}
 0 \le p_1 &&\\ \label{3.4.3b}
-2 p_1 \le  p_2\le p_1 &&\\ \label{3.4.3c}
-\frac{1}{2}\sqrt{p_1(2p_1-p_2)} \le p_3,p_4\le
\frac{1}{2}\sqrt{p_1(2p_1-p_2)} &&\\ \nonumber
\frac{p_1}{2}+\frac{1}{2p_1-p_2}\left(
 2p_3 p_4 -\frac{1}{2}\sqrt{(2p_1^2-p_1 p_2 -4 p_3^2)(2p_1^2-p_1 p_2 -4 p_4^2)}
\right) &\le p_5\le \\   \label{3.4.3e}
\frac{p_1}{2}+\frac{1}{2p_1-p_2}\left(
 2p_3 p_4 +\frac{1}{2}\sqrt{(2p_1^2-p_1 p_2 -4 p_3^2)(2p_1^2-p_1 p_2 -4 p_4^2)}\right)
\end{eqnarray}

Again, it follows by the convex cone property of ${\cal C}_\Phi$
that a system of $N$ dimers with $s=1/2$ has the DGS property if
${\Bbb J}$ can be written as a sum of $4\times 4$-submatrices
satisfying the above inequalities.\\
For another application of the direct method to a homogeneous
ring of $3$ dimers see example \ref{sec4.2}.

\subsection{The classical case}
\label{sec3.4}
In the classical case it is possible to completely characterize
all DGS systems. Recall that $\epsilon_{ij}=\epsilon_{ji}$ denotes
the uniform interaction strength between two dimers according to
(\ref{2.2.3}). For any ${\Bbb J}\in{\cal J}_\Phi^{\small{cl}}$
define an $N\times N$-matrix ${\Bbb G}({\Bbb J})$ with entries
\begin{eqnarray}\label{3.4.1a}
G_{ii}&=&J_{i0,i1} \mbox{ for all }i=1,\ldots,N\;,\\ \label{3.4.1b}
G_{i,j}&=& \epsilon_{ij} \mbox{ for all }i\neq j=1,\ldots,N\;.
\end{eqnarray}
Then we have the following result:
\begin{theorem}
\label{T3}
Let ${\Bbb J}\in{\cal J}_\Phi^{\small{cl}}$, then
${\Bbb J}\in{\cal C}_\Phi^{\small{cl}}$  iff ${\Bbb G}({\Bbb J})$
is positive semi-definite.
\end{theorem}
Recall that ${\Bbb G}({\Bbb J})\ge 0$  iff the $N$ principal minors
$\det(G_{ij})_{i,j=1,\ldots,n}\ge 0$ for $n=1,\ldots,N$.
Hence for classical spin systems the DGS property can be checked
by testing $N$ inequalities. \\

This result is also relevant for quantum spin systems since we have
the following:
\begin{prop}
\label{P6}
 ${\cal C}_\Phi^{\small{cl}}\subset {\cal C}_\Phi^s$ for all
 $s=\frac{1}{2},1,\frac{3}{2},\ldots$.
\end{prop}

\section{Examples}
\subsection{}
\label{sec4.1} 
\begin{figure}
  \includegraphics[width=\columnwidth]{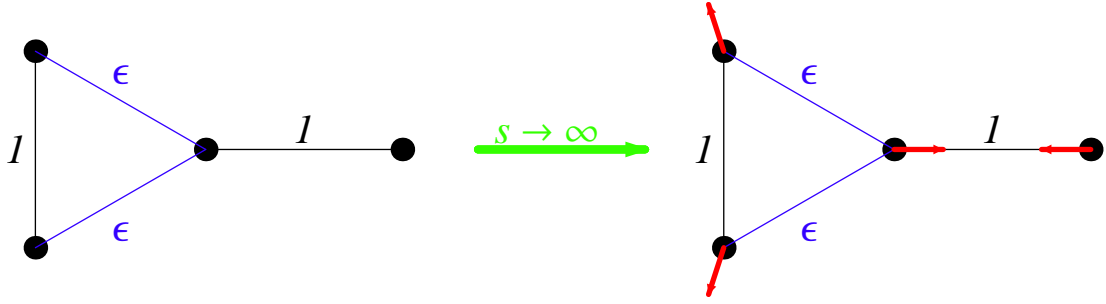}
\caption{\label{fig03}The system at the l.~h.~s.~cannot be a DGS system for
fixed $\epsilon>0$ and arbitrary $s$, since its classical limit at
the r.~h.~s.~is not DGS. The classical ground state is indicated
by small arrows.}
\end{figure}

One of the simplest potential DGS  systems ${\Bbb
J}(\epsilon)$, see figure \ref{fig03}, shows an interesting
effect: For given $s$ and sufficiently small $\epsilon$ it is a
DGS system by virtue of proposition \ref{P3}. But if $\epsilon>0$
is fixed and $s$ increases, it eventually looses the DGS property.
Otherwise we would get a contradiction since ${\Bbb
J}(\epsilon)\notin {\cal C}_\Phi^{\small{cl}}$  by theorem
\ref{T2} and the (normalized) ground state energy must converge
for $s\rightarrow\infty$ towards its classical value as a
consequence of the Berezin/Lieb inequality
\cite{Berezin75}\cite{Lieb73}.

\subsection{}
\label{sec4.2}

\begin{figure}
  \includegraphics[width=10cm]{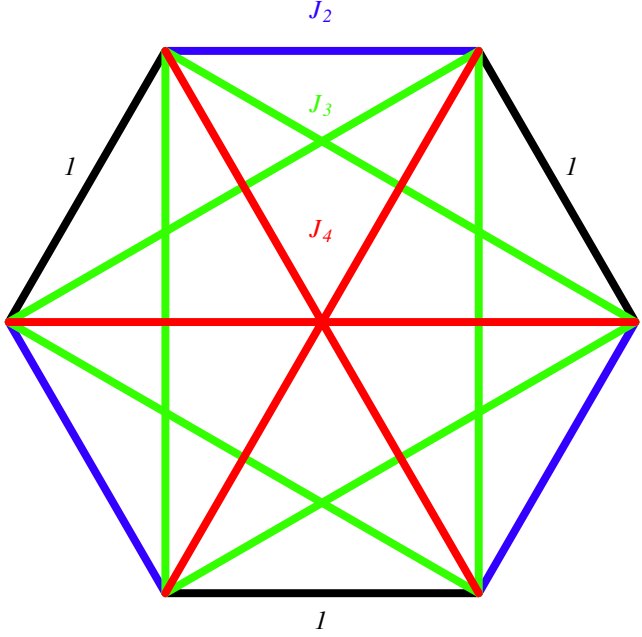}
\caption{\label{fig04}A system of $3$ dimers with special coupling constants
$J_1=1,J_2, J_4$ and $J_3=\frac{1}{2}(J_2+J_4)$.}
\end{figure}

Another $s=1/2$ system for which ${\cal C}_\Phi$ can be directly
calculated by the method sketched in section \ref{sec3.3} consists
of $3$ dimers with equal coupling strength $J_1$ and three further
coupling constants $J_2,J_3,J_4$, see figure \ref{fig04}.
According to theorem \ref{T1} we must have
$J_3=\frac{1}{2}(J_2+J_4)$ if this system has a DGS. Hence only
three independent variables, say, $J_1,J_2$ and $J_4$ are left.
These define a $3$-dimensional subspace of ${\cal J}_\Phi$ and a
corresponding sub-cone ${\cal C}'_\Phi$ of ${\cal C}_\Phi$.
${\cal C}'_\Phi$ can be represented by a convex subset ${\cal K}_\Phi$ of
the $J_2-J_4-$plane which is the intersection of the cone ${\cal
C}'_\Phi$ and the hyperplane $J_1=1$. The permutation
$(12)(34)(56)$ swaps $J_2$ and $J_4$ leaving ${\cal C}'_\Phi$
invariant. Hence ${\cal K}_\Phi$ is symmetric w.~r.~t.~reflections
at the axis $J_2=J_4$.
\\

\begin{figure}
  \includegraphics[width=\columnwidth]{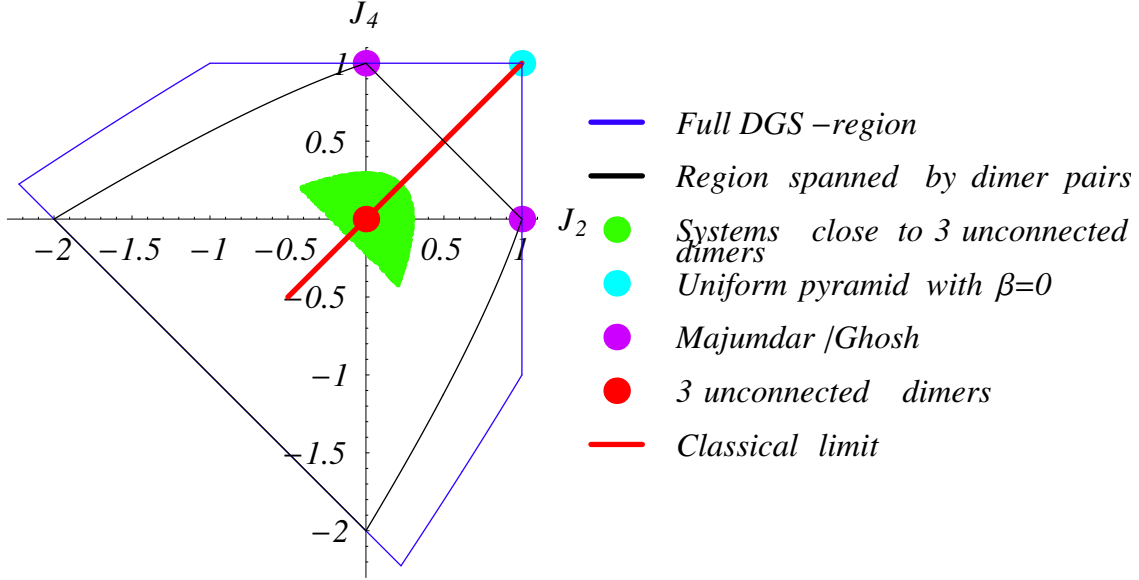}
\caption{\label{fig05}The set ${\cal K}_\Phi$ of points with coordinates
$(J_2,J_4)$ such that the system shown in figure \ref{fig04} is a DGS
system. Certain prominent subsets and points are displayed which
re explained in section \ref{sec4.2}.}
\end{figure}

${\cal K}_\Phi$ is bounded by the lines $J_4=1$, $J_2=1$  and
$J_4+J_2=-2$ and by two curves with small but finite curvature which lie symmetric to the axis
$J_2=J_4$, see figure \ref{fig05}. Prominent points of  ${\cal
K}_\Phi$ are
\begin{itemize}
\item  $(1,1)\;:$ Uniform pyramid with $\beta=0$,
\item $(1.0)$ and $(0,1)\;:$ Majumdar/Ghosh rings,
\item $(0,0)\;:$ Three unconnected dimers.
\end{itemize}
The small neighborhood of $(0,0)$ which belongs to ${\cal K}_\Phi$
according to proposition \ref{P3} is also shown. We see that
the bound in (\ref{3.2.2}) is far from being optimal in this case.
Further we have displayed the convex subset ${\cal L}_\Phi \subset {\cal K}_\Phi$
generated by three dimer pairs according to proposition
\ref{P4}\\
For larger $s>\frac{1}{2}$ the figures ${\cal K}_\Phi^{s}$ would
shrink towards their classical limit which turns out to be the
line segment
\begin{equation}\label{4.2.1}
-\frac{1}{2}\le J_2=J_4 \le 1
\;.
\end{equation}
As mentioned before, the boundary of ${\cal K}_\Phi$ corresponds to a
degeneracy of the ground state. It is easy to identify the ``rival" ground states
for the straight parts of the boundary: If $J_2=1$ the second Majumdar/Ghosh
dimerized state becomes a rival ground state, analogously for $J_4=1$.
If the system approaches the line $J_2+J_4=-2$ the ferromagnetic
ground state $|\uparrow \uparrow\uparrow\uparrow\uparrow\uparrow\rangle$
becomes the rival ground state. The curved parts of the boundary of  ${\cal K}_\Phi$
correspond to continuously varying families of rival ground states.

\section{Proofs}
\label{sec5}
\subsection{Proof of theorem \ref{T1}}
\label{sec5.1}
We rewrite the Hamiltonian (\ref{2.1.1}) in the form
\begin{eqnarray}
\label{5.1.1a}
H({\Bbb J})
&=&
\sum_{\mu\neq\nu}J_{\mu\nu}
\:\bi{s}_\mu\cdot \bi{s}_\mu\\   \label{5.1.1b}
&=&
\sum_{i\neq j}\sum_{\epsilon,\delta=0}^1 J_{i\epsilon,j \delta}
\:\bi{s}_{i\epsilon}\cdot \bi{s}_{j\delta}
+ 2 \sum_{i=1}^N J_{i0,i1} \bi{s}_{i0}\cdot \bi{s}_{i1}  \\     \label{5.1.1c}
&\equiv &
\sum_{i<j}\overline{H}_{ij}
\;,
\end{eqnarray}
where the distribution of the terms of the second sum in
(\ref{5.1.1b}) to the terms $\overline{H}_{ij}$ is arbitrary. We have
$\overline{H}_{ij}=H_{ij}\otimes \Eins^{(ij)}$ such that  $H_{ij}$
acts on ${\cal H}_{ij}={\cal H}_i\otimes{\cal H}_j$ and $\Eins^{(ij)}$
on the remaining factors. Recall that the dimerized state has the form
\begin{equation}\label{5.1.2}
\Phi = \bigotimes_{i=1}^N  [i0,i1]
\;,
\end{equation}
where $[i0,i1]$ denotes the AF dimer ground state (\ref{2.1.3}) in
${\cal H}_i={\cal H}_{i0}\otimes {\cal H}_{i1}$.
\begin{lemma}
\label{L2}
$\Phi$ is an eigenstate of $H({\Bbb J})$ iff
$[i0,i1]\otimes [j0,j1]$ is an eigenstate of $H_{ij}$ for all $i<j=2,\ldots N$.
\end{lemma}
Proof: The if-part is obvious. To prove the only-if-part we
consider an orthonormal basis $(e_\lambda^{(i)})_{\lambda=0,\ldots,L}$
in ${\cal H}_i$ such that $e_0^{(i)}=[i0,i1]$ where
$L=(2s+1)^2-1$. The action of $H_{ij}$ on $[i0,i1]\otimes [j0,j1]$
may be written as
\begin{equation}\label{5.1.3}
H_{ij}(e_0^{(i)}\otimes e_0^{(j)})=
\sum_{\lambda,\mu=0}^L h_{\lambda\mu}^{(ij)} e_\lambda^{(i)}\otimes e_\mu^{(j)}
\;.
\end{equation}

Let $(e_\lambda^{(i)}\otimes e_\mu^{(j)})_\Phi\equiv
e_0^{(1)}\otimes e_0^{(2)}\otimes\ldots e_\lambda^{(i)}\otimes\ldots e_\mu^{(j)}\otimes\ldots\otimes e_0^{(N)}$.
Then
\begin{eqnarray}\label{5.1.4a}
H({\Bbb J})\Phi
&=&
\sum_{i<j}\overline{H}_{ij} (e_0^{(i)}\otimes e_0^{(j)})_\Phi
\\   \label{5.1.4b}
&=&
\sum_{i<j}\sum_{\lambda\mu}  h_{\lambda \mu}^{(ij)} (e_\lambda^{(i)}\otimes e_\mu^{(j)})_\Phi  = E \Phi
=E  \sum_{i<j} (e_0^{(i)}\otimes e_0^{(j)})_\Phi
\;.
\end{eqnarray}
The terms at the l.~h.~s.~of (\ref{5.1.4b}) proportional to
$(e_\lambda^{(i)}\otimes e_\mu^{(j)})_\Phi$ with
$\lambda,\mu=1,\ldots,L$ cannot cancel since they occur only once
in the sum $\sum_{i<j}$. But they don't occur at the r.~h.~s.~of
(\ref{5.1.4b}), hence $h_{\lambda \mu}^{(ij)}=0$ for all
$\lambda,\mu=1,\ldots,L$. $H_{ij}$ maps the subspace of ${\cal
H}_{ij}$ with total spin quantum number $S_{ij}=0$ onto itself,
hence $h_{0\mu}^{(ij)}=h_{\lambda 0}^{(ij)}=0$ for all
$\mu,\lambda=1,\ldots,L$. Thus only $h_{00}^{(ij)}$ may be
non-zero, which means that $e_0^{(i)}\otimes e_0^{(j)}$ is an
eigenstate of $H_{ij}$.
\hspace*{\fill}\rule{3mm}{3mm}  \\

In view of lemma \ref{L2} we only need to consider the case of
$N=2$ dimers with indices $i<j$ in the remaining part of the proof.
We set $\Phi=[i0,i1]\otimes
[j0,j1]$ and rewrite the indices according to
\begin{equation}\label{5.1.5}
 (i0)\equiv 0,\:  (i1)\equiv 1,\: (j0)\equiv 2,\: (j1)\equiv 3
 \;,
\end{equation}
Consider a modified Hamiltonian of the form
\begin{equation}\label{5.1.6}
 H'=\sum_{(\mu\nu)\in\{02,03,12,13\}}J_{\mu\nu} \bi{s}_\mu \cdot \bi{s}_\nu
 \;.
\end{equation}
Obviously, $\Phi$ is an eigenvector of $H_{ij}$ iff it is an
eigenvector of $H'$, since the difference between $H_{ij}$ and
$H'$ consists of two dimer Hamiltonians. We note that for fixed
$\mu\in\{0,1,2,3\}$ the three operators $\bi{s}_\mu^{(i)},
i=1,2,3$ form an ``irreducible tensor operator", i.~e.~they span a
$3$-dimensional irreducible subspace with quantum number $S=1$.
Here and henceforward ``irreducible" will always be understood as
``irreducible w.~r.~t.~the product representation of $SU(2)$ in
${\cal H}_{ij}$ (or similar spaces)". In order to apply the
Wigner-Eckhardt theorem (WE), see for example \cite{BL81}, we
consider states of the form
\begin{equation}\label{5.1.7}
\psi =\psi_{01}\otimes \psi_{23}
\;,
\end{equation}
such that $\psi_{01}\in{\cal H}_{01}$ (resp.~$\psi_{23}\in{\cal H}_{23}$)
belong to irreducible subspaces of ${\cal H}_{01}$ (resp.~${\cal H}_{23}$)
characterized by their dimension $2S_{01}+1$  (resp.~$2S_{23}+1$).
Recall that WE yields ``selection rules" of the following kind: The matrix element
of a component of a tensor operator with representation $D$ between two states belonging to irreducible
representations $D_1$ and $D_2$ is nonzero only if $D$
is contained in the product representation of $D_1$ and $D_2$. In the case of irreducible $SU(2)$
representations which are characterized by quantum numbers, say, $S$, $S_1$ and $S_2$,
the above condition simply reads: $|S_1-S_2|\le S \le S_1+S_2$.
The dimer ground states $[01]$ and $[23]$ of course span $S_2=0$ representations.
Then WE yields:
\begin{lemma}
\label{L3}
$\langle \psi|H'|\Phi\rangle \neq 0$
only if $S_{01}=S_{23}=1$.
\end{lemma}
Proof: It will suffice to consider only one term of $H'$, for example
$J_{02}\bi{s}_0^{(i)} \otimes \bi{s}_2^{(i)}$, since analogous arguments
apply for the other terms. We skip the factor $J_{02}$ and write
\begin{eqnarray}\label{5.1.8a}
\langle \psi|\bi{s}_0^{(i)} \otimes \bi{s}_2^{(i)}|\Phi\rangle
&=&
\langle \psi_{01}\otimes \psi_{23}|\bi{s}_0^{(i)} \otimes \bi{s}_2^{(i)}|[01]\otimes[23]\rangle
\\  \label{5.1.8b}
&=&
\langle \psi_{01}|\bi{s}_0^{(i)} |[01]\rangle
\langle  \psi_{23}| \bi{s}_2^{(i)}|[23]\rangle
\;.
\end{eqnarray}
The first factor vanishes by WE if $S_{01}\neq 1$, the second one if $S_{23}\neq 1$.
\hspace*{\fill}\rule{3mm}{3mm}  \\
Especially, $\langle \Phi|H'|\Phi\rangle=0$ and hence, if $\Phi$
is an eigenvector of $H'$ the corresponding eigenvalue can only be
zero.\\
It is well-known that the irreducible subspaces of ${\cal H}_{01}$
are eigenspaces of the permutation $\pi_{01}$ with eigenvalues
$(-1)^{S_{01}+2s}$, analogously for $\pi_{23}$. For example, if
$s=1/2$, the $S_{01}=0$ singlet subspace of ${\cal H}_{01}$ is
spanned by the antisymmetric state
$\frac{1}{\sqrt{2}}(\uparrow\downarrow-\downarrow\uparrow)$,
whereas the $S_{01}=1$ triplet subspace is symmetric. The terms
$\bi{s}_\mu\cdot\bi{s}_\nu$ occurring in $H'$ can be generated
from $\bi{s}_0\cdot\bi{s}_2$ by applying suitable permutations
$\pi_{01}$ and $\pi_{23}$. Hence, using lemma \ref{L3} and the
above-mentioned symmetry of $\psi$ under permutations, we obtain
\begin{eqnarray}\label{5.1.9a}
\langle \psi|H' |\Phi \rangle
&=&
\sum_{(\mu\nu)\in\{02,03,12,13\}}J_{\mu\nu} \langle \psi|\bi{s}_\mu \cdot \bi{s}_\nu |\Phi\rangle
\\ \label{5.1.9b}
&=&
(J_{02}+J_{13}-J_{03}-J_{12})
\langle \psi|\bi{s}_0\cdot \bi{s}_2 |\Phi\rangle
\;.
\end{eqnarray}
Since $\langle \psi|\bi{s}_0 \cdot \bi{s}_2 |\Phi\rangle\neq 0$ for a
suitable $\psi$ we conclude that  $J_{02}+J_{13}-J_{03}-J_{12}=0$
iff $H'\Phi=0$  iff $\Phi$ is an eigenvector of $H'$. Together
with lemma \ref{L2} this concludes the proof of theorem \ref{T1}.

\subsection{Proof of proposition \ref{P1}}
\label{sec5.2}
Using the same notation as in section \ref{sec5.1}
we conclude by WE that
\begin{eqnarray}\label{5.2.1}
\langle [01]\otimes \psi_{23}\left|\bi{s}_\mu^{(i)}\otimes\bi{s}_\nu^{(i)}\right|[01]\otimes \psi_{23}\rangle
&=&
\langle [01]\left|\bi{s}_\mu^{(i)}\right|[01] \rangle
\langle  \psi_{23}\left|\bi{s}_\nu^{(i)}\right|\psi_{23}\rangle
=0
\end{eqnarray}
for all $(\mu\nu)\in\{02,03,12,13 \}$. For fixed $j\in\{1,\ldots,N\}$ let
\begin{equation}\label{5.2.2}
\Phi'=\bigotimes_{\stackrel{\scriptstyle i=1}{(i\neq j)}}^N [i0,i1] \otimes |s,s\rangle
\;,
\end{equation}
where $|s,s\rangle$ is the ferromagnetic ground state in ${\cal H }_j$.
By (\ref{5.2.1}) and analogous equations for permuted indices the expectation value
$\langle \Phi'\left| H({\Bbb J})\right|\Phi' \rangle$
contains no interaction terms between dimers and thus
\begin{eqnarray}\label{5.2.3a}
\langle \Phi'\left| H({\Bbb J})\right|\Phi' \rangle
&=&
2\left[
-\sum_{\stackrel{\scriptstyle i=1}{(i\neq j)}}^N
J_{i0,i1}s(s+1) + J_{j0,j1} s^2
\right]  \\ \label{5.2.3b}
&&
\ge
\langle \Phi\left| H({\Bbb J})\right|\Phi \rangle
=
2\left[
-\sum_{i=1}^N
J_{i0,i1}s(s+1)
\right]
\;.
\end{eqnarray}
In (\ref{5.2.3b}) we used the assumption of proposition \ref{P1}
that $\Phi$ is a ground state of $H({\Bbb J})$. It follows that
$J_{j0,j1}s^2 \ge - J_{j0,j1} s(s+1)$ and hence $J_{j0,j1}\ge 0$,
which concludes the proof.

\subsection{Proof of theorem \ref{T2}}
\label{sec5.2}
First we want to show that (\ref{2.2.1}) also holds for classical DGS systems.
\begin{lemma}
\label{L4}
Let ${\Bbb J}\in {\cal C}_\Phi^{\small{cl}}$. Then
\begin{equation}\label{5.2.4}
\widetilde{J}_{ij}\equiv J_{i0,j0}+J_{i1,j1}-J_{i0,j0}-J_{i1,j0}=0
\end{equation}
for all $i<j=2,\ldots,N$.
\end{lemma}
Proof: By assumption, any classical state $\bi{s}$ satisfying
\begin{equation}\label{5.2.5}
\bi{s}_{i0}+\bi{s}_{i1}=\bi{0} \mbox{ for all } i=1,\ldots,N
\end{equation}
minimizes the energy $H({\Bbb J},\bi{s})$. For such states we may write
\begin{eqnarray}
\label{5.2.5a}
H({\Bbb J},\bi{s})
&=&
\sum_{\mu\neq \nu} J_{\mu\nu}\bi{s}_\mu\cdot\bi{s}_\nu
\\   \label{5.2.5b}
&=&
2\sum_{1\le i<j\le N} \sum_{\epsilon,\delta=0}^1
J_{i\epsilon,j\delta}\;\bi{s}_{i\epsilon}\cdot\bi{s}_{j\delta}
-2\sum_{i=1}^N J_{i0,i1}
\\   \nonumber
&=&
2\sum_{1\le i<j\le N}
\left(
J_{i0,j0}+ J_{i1,j1}-J_{i0,j1}-J_{i1,j0}
\right) \bi{s}_{i0}\cdot\bi{s}_{j0}
\\ \label{5.2.5c}
&&
-2\sum_{i=1}^N J_{i0,i1}
\;.
\end{eqnarray}
The function $\bi{s}\mapsto H({\Bbb J},\bi{s})$ is constant for all $\bi{s}\in{\cal P}$
satisfying (\ref{5.2.5}). The above equations show that also the function
$(\bi{s}_i)_{i=1,\ldots,N} \mapsto \widetilde{H}(\bi{s})\equiv\sum_{1\le i<j\le N}  \widetilde{J}_{ij}
\bi{s}_i\cdot \bi{s}_j$ is constant. We write
\begin{equation}\label{5.2.6}
\widetilde{H}(\bi{s})
=
\bi{s}_1\cdot\sum_{2\le j\le N}\widetilde{J}_{1j}\bi{s}_j
+ \sum_{2\le i<j\le N}   \widetilde{J}_{ij}\bi{s}_i \cdot  \bi{s}_j
\;.
\end{equation}
Since $\widetilde{H}(\bi{s})$ is independent of $\bi{s}_1$, the second factor
in the first scalar product in (\ref{5.2.6}) must vanish: $\sum_{2\le j\le N}\widetilde{J}_{1j}\bi{s}_j=0$.
By choosing $\bi{s}_j\perp \bi{s}_2$ for all $j>2$ we conclude $\widetilde{J}_{12}=0$.
This concludes the proof of the proposition since the numbering of the dimers is arbitrary.
\hspace*{\fill}\rule{3mm}{3mm}  \\
Next we want to show (\ref{2.2.3}):
\begin{lemma}
\label{L5}
If ${\Bbb J}\in {\cal C}_\Phi^{\small{cl}}$  then
\begin{equation}\label{5.2.7}
J_{i0,j0}= J_{i0,j1}=J_{i1,j0}=J_{i1,j1}\equiv\epsilon_{ij}
\end{equation}
\end{lemma}
Proof: It will suffice to show $J_{i0,j1}=J_{i1,j0}$, since  $J_{i0,j0}=J_{i1,j1}$
follows by applying the permutation $(01)$ and the remaining identity $J_{i0,j1}=J_{i0,j0}$
by (\ref{5.2.4}).\\
Consider the state $\bi{s}(\alpha)$ defined by
\begin{eqnarray}\label{5.2.8a}
\bi{s}_{i0}
&=&
{\left(\begin{array}{r}1\\0\\0 \end{array}\right)},
\bi{s}_{i1}= {\left(\begin{array}{r}-\cos\alpha\\ \sin\alpha\\0 \end{array}\right)},
\bi{s}_{j0}= {\left(\begin{array}{r}0\\1\\0 \end{array}\right)},\;
\bi{s}_{j1}= {\left(\begin{array}{r}-\sin\alpha\\ -\cos\alpha\\0 \end{array}\right)}
\\ \nonumber
\mbox{and} &&\\
\label{5.2.8b}
\bi{s}_{k0}
&=&
- \bi{s}_{k1} = {\left(\begin{array}{r}0\\0\\1 \end{array}\right)}
\mbox{ for all } k\neq i,j.
\end{eqnarray}
Here $\alpha$ is an arbitrary angle $0\le\alpha<\pi$ to be fixed later.
Let $E(\alpha)$ denote the energy of this state. We conclude
\begin{eqnarray}
\nonumber
\frac{1}{2}E(\alpha)
&=&
J_{i0,i1}\bi{s}_{i0}\cdot\bi{s}_{i1}+
J_{j0,j1}\bi{s}_{j0}\cdot\bi{s}_{j1}+
J_{i0,j0}\bi{s}_{i0}\cdot\bi{s}_{j0}+
J_{i1,j1}\bi{s}_{i1}\cdot\bi{s}_{j1}
\\
\label{5.2.9a}
&&
+J_{i0,j1}\bi{s}_{i0}\cdot\bi{s}_{j1}+
J_{i1,j0}\bi{s}_{i1}\cdot\bi{s}_{j0}
-\sum_{\stackrel{\scriptstyle k=1}{(k\neq i,j)}} J_{k0,k1}
\\
\label{5.2.9b}
&=&
-\left(J_{i0,i1}+J_{j0,j1}\right) \cos\alpha
+ \left(J_{i1,j0}-J_{i0,j1}\right) \sin\alpha
-\sum_{\stackrel{\scriptstyle k=1}{(k\neq i,j)}} J_{k0,k1}
\\
\label{5.2.9c}
&=&
\left(J_{i0,i1}+J_{j0,j1}\right) (1-\cos\alpha)
+ \left(J_{i1,j0}-J_{i0,j1}\right) \sin\alpha
+\frac{1}{2}E(0)\;.
\end{eqnarray}
It is obvious that the state $\bi{s}(\alpha)$ defined by
(\ref{5.2.8a}) and (\ref{5.2.8b}) is a classical DGS for
$\alpha=0$, hence $E(0)$ is the ground state energy. If
$J_{i0,i1}+J_{j0,j1}\le 0$ and $J_{i1,j0}-J_{i0,j1}\neq 0$ we may
choose the sign of $\alpha$ such that $E(\alpha)<E(0)$ which is
impossible due to the last statement. Thus we may assume
$J_{i0,i1}+J_{j0,j1}>0$. Hence the expression (\ref{5.2.9c})
has its minimum at a value $\alpha=\alpha_0$  defined by
\begin{equation}\label{5.2.10}
\tan\alpha_0
=
-\frac{J_{i1,j0}-J_{i0,j1}}{J_{i0,i1}+J_{j0,j1}}
\;.
\end{equation}
After some algebra we obtain for the corresponding energy
\begin{equation}\label{5.2.11}
\frac{1}{2}E(\alpha_0)=
(J_{i0,i1}+J_{j0,j1})
\left(
1-\sqrt{1+\left(\frac{J_{i1,j0}-J_{i0,j1}}{J_{i0,i1}+J_{j0,j1}}\right)^2}
\right)
+ \frac{1}{2}E(0)
\;,
\end{equation}
which is less than $\frac{1}{2}E(0)$ if not  $J_{i1,j0}-J_{i0,j1}=0$.
\hspace*{\fill}\rule{3mm}{3mm}  \\
It remains to show (\ref{2.2.4}):
\begin{lemma}
\label{L6}
If ${\Bbb J}\in {\cal C}_\Phi^{\small{cl}}$  then
\begin{equation}\label{5.2.12}
J_{i0,i1}\ge 0
\end{equation}
for all $i=1,\ldots,N$.
\end{lemma}
Proof: We rewrite $H({\Bbb J},\bi{s})$ using (\ref{5.2.7}),
$\mu_i \equiv J_{i0,i1}$ and $\bi{S}_i\equiv\bi{s}_{i0}+\bi{s}_{i1}$,
$\bi{S}_{ij}\equiv\bi{S}_i+\bi{S}_j$ for $i<j=2,\ldots,N$:
\begin{eqnarray}\nonumber
H({\Bbb J},\bi{s})
&=&
2 \sum_{i=1}^N \mu_i  \bi{s}_{i0}\cdot\bi{s}_{i1}
\\ \label{5.2.13a}
&+&
2\sum_{1\le i<j\le N} \epsilon_{ij}
(\bi{s}_{i0}\cdot\bi{s}_{j0}+\bi{s}_{i1}\cdot\bi{s}_{j0}+\bi{s}_{i0}\cdot\bi{s}_{j1}+\bi{s}_{i1}\cdot\bi{s}_{j1})
\\
\label{5.2.13b}
&=&
\sum_{i=1}^N \mu_i (\bi{S}_i^2-2)+\sum_{1\le i\neq j\le N} \epsilon_{ij} \bi{S}_i\cdot\bi{S}_j
\;.
\end{eqnarray}
The energy of the DGS will be $E_0=-2\sum_{i=1}^N \mu_i$.
Assume that for some $i\in\{1,\ldots,N\}$ we have $\mu_i<0$ and
consider a state $\bi{s}$ such that $\bi{S}_j=\bi{0}$ for all $j\neq i$ and
$|\bi{S}_i|=2$. For this state we obtain $H({\Bbb J},\bi{s}) =4 \mu_i+E_0$ which contradicts
the assumption that the system has a DGS.
\hspace*{\fill}\rule{3mm}{3mm}  \\
This completes the proof of theorem \ref{T2}.

\subsection{Proof of theorem \ref{T3}}
\label{sec5.3}
In order to prove theorem \ref{T3} we stick to the notation of the last section and
rewrite the energy of a classical
DGS system  (\ref{5.2.13b}) in the form
\begin{equation}\label{5.3.1}
E(\vec{S})=\langle \vec{S}|{\Bbb M}|\vec{S}\rangle
-2 \sum_{i=1}^N \mu_i
\;.
\end{equation}
Here $\vec{S}$ denotes a $3N$-dimensional vector with components
\begin{equation}\label{5.3.2}
\vec{S}=(S_1^{(1)},S_1^{(2)},S_1^{(3)},\ldots,S_N^{(1)},S_N^{(2)},S_N^{(3)})\;,
\end{equation}
and ${\Bbb M}\equiv {\Bbb G}\otimes \Eins_{\:_3}$, where $\Eins_{\:3}$ is the $3\times 3$ unit
matrix and ${\Bbb G}$ has the components
\begin{eqnarray}\label{5.3.3a}
G_{ii}&=&\mu_i,\; i=1,\ldots,N\\  \label{5.3.3b}
G_{ij}&=&\epsilon_{ij},\;1\le i\neq j\le N\;.
\end{eqnarray}
We have to show that the system is DGS iff ${\Bbb M}$, or,
equivalently, ${\Bbb G}$ is positive semi-definite. This follows
immediately from ${\Bbb M}\ge 0$ iff $E(\vec{S})\ge -2\sum_{i=1}^N
\mu_i = E_0$ for all $\vec{S}$ such that $|\bi{S}_i|\le 2$ for all
$i=1,\ldots,N$.

\subsection{Proof of proposition \ref{P2}}
\label{sec5.4}
Due to (\ref{3.1.1a}) and (\ref{3.1.1b})  the Hamiltonian of a uniform pyramid
can be written as
\begin{equation}\label{5.4.1}
H({\Bbb J})=(\alpha-\beta)(\bi{S}_N^2-2Ns(s+1))+\beta \left[ (\bi{S}_N+\bi{s}_{N+1,0})^2
-(2N+1)s(s+1)\right]
,
\end{equation}
where
\begin{equation}\label{5.4.2}
\bi{S}_N\equiv \sum_{i=1}^N \sum_{\delta=0}^1 \bi{s}_{i\delta}
\;.
\end{equation}
The eigenvalues of $\bi{S}_N^2$ are of the form $S(S+1)$, where $S=0,1,\ldots,2Ns$.
Obviously, the choice $S=0$ minimizes (\ref{5.4.1}) if $\beta$ is small enough.
In this case $\Phi$ is a ground state since it has $S=0$. If $\beta$ increases,
it will eventually reach a certain value $\beta_0$ where $S=1$ states have the
same energy as $\Phi$. The other values $S>1$ can be excluded. For $\beta=\beta_0$
and $s>1/2$ the coincidence of the energies implies
\begin{eqnarray}\nonumber
(\alpha-\beta_0)(-2Ns(s+1))
&+&
\beta_0(s(s+1)-(2N+1)s(s+1))\\
\label{5.4.3a}
=(\alpha-\beta_0)(1\cdot 2-2Ns(s+1))
&+&
\beta_0((s-1)s)-(2N+1)s(s+1)
\;,
\end{eqnarray}
hence
\begin{equation}\label{5.4.4}
\beta_0=\frac{\alpha}{1+s}
\end{equation}
and the system has a DGS for $\beta\le\beta_0$.
The value $\beta_0=\alpha$ in the case $s=1/2$ follows analogously.
The only difference is that in this case the minimal eigenvalue of
$(\bi{S}_N+\bi{s}_{N+1,0})^2$ will be $s(s+1)=3/4$ and not $(s-1)s$ as in
the case $s>1/2$.

\subsection{Proof of proposition \ref{P3}}
\label{sec5.5}
We consider the matrix $\stackrel{\circ}{\Bbb J}$ of a system
of $N$ uncoupled AF dimers, i.~e.~
\begin{equation}\label{5.5.1}
\lambda_i\equiv\stackrel{\circ}{J}_{i0,i1} > 0 \mbox{ for all }
i=1,\ldots,N \mbox{ and } \stackrel{\circ}{J}_{\mu\nu}=0 \mbox{ else.}
\end{equation}
Without loss of generality we may assume
\begin{equation}\label{5.5.2}
\lambda_1=\lambda\equiv\min\{ \lambda_i|i=1\ldots N\}
\;.
\end{equation}
The other assumptions are as in proposition \ref{P3}. Let
$\delta_{\small{min}}$ denote the lowest eigenvalue of $\Delta$, hence
${\cal x}(\Delta)=|\delta_{\small{min}}|$.  Consider any $\Psi\in{\cal H}$
such that $\|\Psi\|=1$ and $\Psi\perp\Phi$. The two lowest eigenvalues of
$H(\stackrel{\circ}{\Bbb J})$ are
\begin{eqnarray}\label{5.5.3a}
E_0
&=&
-2s(s+1)\sum_{i=1}^N \lambda_i   \quad \mbox{and}
\\  \label{5.5.3b}
E_1
&=&
-2s(s+1)\sum_{i=2}^N \lambda_i  +\lambda_1(2-2s(s+1))
\\     \label{5.5.3c}
&=&
E_0+2\lambda_1
\;.
\end{eqnarray}
$E_0$ belongs to the non-degenerate eigenvector $\Phi$, hence
\begin{equation}\label{5.5.4}
E_1\le \langle \Psi| H(\stackrel{\circ}{\Bbb J})|\Psi\rangle
\;.
\end{equation}
The expectation value of the Hamiltonian $H(\Delta)$ can be estimated
as follows
\begin{eqnarray}\label{5.5.5a}
\langle \Psi|H(\Delta)|\Psi\rangle
&=&
\sum_{\mu\nu}\Delta_{\mu\nu} \langle \Psi|\bi{s}_\mu\cdot\bi{s}_\nu|\Psi\rangle
\\   \label{5.5.5b}
&\ge&
\delta_{\small{min}} \sum_{\mu}  \langle \Psi|\bi{s}_\mu^2|\Psi\rangle
=\delta_{\small{min}} N s(s+1)
\\   \label{5.5.5c}
&\ge&
-2\lambda_1
\;,
\end{eqnarray}
using (\ref{3.2.2}) and $\delta_{\small{min}}<0$ since $\Tr(\Delta)=0$.
Thus
\begin{eqnarray}\label{5.5.6a}
\langle \Psi|H({\Bbb J})|\Psi\rangle
&=&
\langle \Psi|H(\stackrel{\circ}{\Bbb J})|\Psi\rangle  +
\langle \Psi|H(\Delta)|\Psi\rangle
\\   \label{5.5.6b}
&\ge&
E_1-2\lambda_1=E_0
\;,
\end{eqnarray}
using (\ref{5.5.4}), (\ref{5.5.5c}) and (\ref{5.5.3c}).
This proves that $\Phi$ is a ground state of  $H({\Bbb J})$.

\subsection{Proof of proposition \ref{P6}   }
\label{sec5.6}
Let ${\Bbb J}\in{\cal C}_\Phi^{\small{cl}}$
and $\Psi\in{\cal H},\; \|\Psi\|=1$. Then
\begin{eqnarray}\label{5.6.1a}
\langle \Psi|H({\Bbb J})|\Psi\rangle
&=&
\sum_{i=1}^N G_{ii}(\langle \Psi|\bi{S}_i^2|\Psi\rangle  -2)+
\sum_{1\le i\neq j\le N} G_{ij}(\langle \Psi|\bi{S}_i\cdot\bi{S}_j|\Psi\rangle
\\   \label{5.6.1b}
&=&
\Tr {\Bbb G}{\Bbb W}+E_0
\;,
\end{eqnarray}
where ${\Bbb W}$ is an $N\times N$- matrix with elements
\begin{equation}\label{5.6.2}
W_{ij}= \langle \Psi|\bi{S}_i\cdot\bi{S}_j|\Psi\rangle
\mbox{ and }
E_0=  \langle \Phi|H({\Bbb J})|\Phi\rangle
\;.
\end{equation}
It can be easily shown that ${\Bbb W}\ge 0$. Since also ${\Bbb
G}\ge 0$ by theorem \ref{T3} we may conclude that $\Tr ({\Bbb
G}{\Bbb W})\ge 0,\;\langle \Psi|H({\Bbb J})|\Psi\rangle \ge E_0$
and hence ${\Bbb J}\in{\cal C}_\Phi^s$ for all
$s=1/2,1,3/2,\ldots$.

\section{Geometrical structure of ${\cal C}_\Phi$}
\label{sec6}
The defining inequalities of ${\Bbb J}\in{\cal C}_\Phi$ are
 \begin{equation}\label{6.1}
\langle \Phi|H({\Bbb J})|\Phi\rangle
\le
\langle \psi|H({\Bbb J})|\psi\rangle
\mbox{ for all }\psi\in{\cal H},\;\|\psi\|=1
\;.
\end{equation}
For fixed $\psi,\;\psi\neq\Phi,$ (\ref{6.1}) defines a closed half space of
the real linear space ${\cal J}_\Phi$.
Thus ${\cal C}_\Phi$ is an intersection of closed half spaces and hence a
{\it closed convex cone}. For the notion of a {\it cone} and related notions, see,
for example, \cite{Schaefer71}.\\
Moreover,  ${\cal C}_\Phi$ is a {\it proper} cone,
i.~e.~${\Bbb J},-{\Bbb J}\in{\cal C}_\Phi$ implies ${\Bbb J}=0$.
Indeed, ${\Bbb J},-{\Bbb J}\in{\cal C}_\Phi$ means that
$\langle \Phi|H({\Bbb J})|\Phi\rangle$ is simultaneously the lowest
and the highest eigenvalue of $H({\Bbb J})$, which, due to $\Tr H({\Bbb J})=0$,
implies $H({\Bbb J})=0$ and ${\Bbb J}=0$.\\
The set of differences ${\cal C}_\Phi-{\cal C}_\Phi$ is a linear subspace of
${\cal J}_\Phi$; by virtue of proposition \ref{P3} ${\cal C}_\Phi$
contains interior points and hence  ${\cal C}_\Phi-{\cal C}_\Phi={\cal J}_\Phi$.
In other words, ${\cal J}_\Phi$ is a {\it generating} cone.\\

A {\it face} $F$ is a convex subset of a convex set ${\cal C}$
such that $\lambda c_1 +(1-\lambda)c_2\in F,\; c_1,c_2\in {\cal
C},\; 0<\lambda<1$ implies $c_1\in F$ and $c_2\in F$. In words: If
a point of $F$ lies in the interior of a segment contained in
${\cal C}$, then the endpoints of that segment will also lie in
$F$. A {\it proper} face is a face different from $\emptyset$ and
${\cal C}$. Special faces are the singletons $F=\{f\}$, such that
$f$ never lies in the interior of a segment contained in ${\cal
C}$; such $f$ are called {\it extremal points} of ${\cal C}$. The
{\it dimension} $d$ of a face $F$ is defined as the dimension of
the affine subspace generated by $F$. Hence extremal points can be
viewed as $0$-dimensional faces. The above definitions follow, for
example, \cite{NariciB85}; other authors reserve the notion of a
``face" to the intersection of ${\cal C}$ with a supporting
hyperplane. By virtue of theorem \ref{T4} both notions coincide
for the faces of ${\cal C}_\Phi$. The boundary of a closed convex
set ${\cal C}$ in a finite-dimensional linear space is the union
of its faces $F$, excluding  $F={\cal C}$. The set of all faces of
${\cal C}$ will form a {\it lattice} w.~r.~t.~the set-theoretic
inclusion of faces, such that $F_1\wedge F_2=F_1\cap F_2$, but
$F_1\vee F_2$ will be the smallest face containing $F_1$ and
$F_2$, which in general larger than $F_1\cup F_2$.
\\

It is possible to more closely characterize the faces of ${\cal
C}_\Phi$. To this end we note that the definition (\ref{6.1}) of
${\cal C}_\Phi$ does not make use of the special form of $\Phi$ as
a product of dimer ground states. Hence ${\cal C}_\phi$ may
defined for arbitrary normalized states  $\phi\in{\cal H}$. We
will, however, always restrict this definition to ${\Bbb J}\in
{\cal J}_\Phi$ such that ${\cal C}_\phi$ will be a closed proper
convex cone in ${\cal J}_\Phi$ also for the general case.
\\
\begin{theorem}
\label{T4}
All faces $F$ of  ${\cal C}_\Phi$ are of the form
\begin{equation}\label{6.2}
F=\bigcap_{i=0}^k {\cal C}_{\phi_i}
\;,
\end{equation}
where $\phi_0=\Phi,\; \phi_i\in {\cal H},$ and $\langle \phi_i,\phi_j\rangle =\delta_{ij}$
for all $i,j=0,\ldots,k$.  \\
Conversely, any intersection of the form (\ref{6.2}) will be a, possibly empty, face of ${\cal C}_\Phi$.
\end{theorem}
The special case $F={\cal C}_\Phi$ is included in (\ref{6.2}) with $k=0$. If ${\Bbb J}$ lies at the
boundary of ${\cal C}_\Phi$, it is contained in a proper face of ${\cal C}_\Phi$ and hence there will be at least
one further ground state different from $\Phi$. Hence we have the following:
\begin{cor}
\label{C2}
${\Bbb J}$ is an interior point of  ${\cal C}_\Phi$ iff $\Phi$ is a non-degenerate
ground state of $H({\Bbb J})$.
\end{cor}
Proof of theorem \ref{T4}:
\begin{enumerate}
\item
Let $F=\bigcap_{i=0}^k {\cal C}_{\phi_i}$  be given with the properties stated in the theorem.
We want to show that $F$ is a face. Clearly, $F$ is a convex subset of ${\cal C}_\Phi$.
Assume
${\Bbb J}=\lambda {\Bbb J}_1 +(1-\lambda){\Bbb J}_2\in F,\; {\Bbb J}_i\in{\cal C}_\Phi,\;0<\lambda<1$.
If $k=0$ we are done, hence we may assume $k>0$.
Let $E_i\equiv \langle \Phi |H({\Bbb J}_i)|\Phi\rangle$ denote the ground state energies and
$E'_i\equiv \langle \phi_1 |H({\Bbb J}_i)|\phi_1\rangle$. By assumption,
\begin{eqnarray}\label{6.3a}
\lambda E'_1+(1-\lambda)E'_2
=
\langle \phi_1|H({\Bbb J})|\phi_1\rangle
&=&
\langle \Phi|H({\Bbb J})|\Phi\rangle \\   \label{6.3b}
&=&
\lambda E_1+(1-\lambda)E_2,  \\   \nonumber
\mbox{hence} && \\ \label{6.3c}
\lambda (E'_1-E_1)+(1-\lambda)(E'_2-E_2)
&=&0
\;.
\end{eqnarray}
In (\ref{6.3c}) both terms are non-negative since $E_1$ and $E_2$
are ground state energies. Hence $E'_1=E_1$ and $\phi_1$ is a
ground state of $H({\Bbb J}_1)$. The same holds for all $\phi_i,\;
i=2,\ldots,k$. This means that ${\Bbb J}_1\in F$, analogously
${\Bbb J}_2\in F$, and thus $F$ is a face of ${\cal C}_\Phi$.
During this proof we will call faces of this kind {\it standard} faces.
\item
We will prove the following:
\begin{lemma}
\label{L7}
Each boundary point of a standard face of ${\cal C}_\Phi$ is contained in a smaller standard face.
\end{lemma}
Proof: Let $F$ be a standard face of the form (\ref{6.2})
and $L$ a line in ${\cal J}_\Phi$ such that
\begin{equation}\label{6.4}
F\cap L =\{ \lambda {\Bbb J}_1+(1-\lambda) {\Bbb J}_2 | 0\le \lambda \le 1\}
,\; {\Bbb J}_1\neq {\Bbb J}_2\;.
\end{equation}
Obviously, ${\Bbb J}_1$ and ${\Bbb J}_2$ are boundary points,
and every boundary point of $F$ can be obtained in this way.
We consider the $1$-parameter analytic family of Hamiltonians
$H(\lambda)=H(\lambda {\Bbb J}_1+(1-\lambda) {\Bbb J}_2)$. There exists
a complete set of
eigenvectors $\psi_j(\lambda)$ of $H(\lambda)$ such that the corresponding
eigenvalues $E_j(\lambda)$ are analytical functions in some neighborhood of
$\lambda=0$, see, for example, \cite{Lan69} theorem 7.10.1. We may
arrange the indices such that the first $k+1$ eigenfunctions
$\psi_0(\lambda),\ldots,\psi_k(\lambda)$ are degenerate and
span the same eigenspace of $H(\lambda)$ as the $\phi_i,\; i=0,\ldots,k$ for all
$\lambda\in[0,1]$. According to the cited theorem they must also be degenerate for
$\lambda\in(-\epsilon,0)$ for some small $\epsilon>0$ since an
analytical function is constant on its total domain of definition if it is locally constant.
But for $\lambda\in(-\epsilon,0)$
the ground states of $H(\lambda)$ must be other eigenvectors, say
$\psi_{k+1}(\lambda),\ldots,\psi_{\ell}(\lambda)$, since
$\lambda {\Bbb J}_1+(1-\lambda) {\Bbb J}_2 \notin F$. By continuity,
$\psi_{k+1}(0),\ldots,\psi_{\ell}(0)$ are still
ground states of $H(0)$. We thus conclude ${\Bbb J}_2\in \bigcap_{i=0}^{\ell}{\cal C}_{\phi_i}$
if $\phi_{k+1}\equiv \psi_{k+1}(0),\ldots,\phi_{\ell}\equiv \psi_{\ell}(0)$.
This proves lemma \ref{L7}.
\item
It remains to show that each face $F$ of  ${\cal C}_\Phi$ is a standard face.
First consider the case where $F$ does not consist of a single extremal point. Let
${\Bbb J}\in F$ be an interior point of $F$ and $(\phi_i)_{i=0,\ldots,k}$ an
orthonormal basis of the eigenspace of $H({\Bbb J})$ corresponding to its lowest
eigenvalue $E_0$. We may set $\phi_0=\Phi$. Let ${\Bbb J}_1\in F$ be arbitrary, but ${\Bbb J}_1\neq {\Bbb J}$
and consider $H(\lambda)=H(\lambda{\Bbb J}+(1-\lambda){\Bbb J}_1),\;\lambda\in{\Bbb R},\;$ as well as
the affine functions $\lambda\mapsto\langle \phi_i|H(\lambda)|\phi_i\rangle,\;i=0,\ldots,k$.
They coincide at $\lambda=1$ since
$\langle \phi_i|H(1)|\phi_i\rangle=\langle \phi_i|H({\Bbb J})|\phi_i\rangle=E_0$.
For some neighborhood of $\lambda=1$,
$\lambda{\Bbb J}+(1-\lambda){\Bbb J}_1\in {\cal C}_\Phi$, since ${\Bbb J}$ is an interior point of $F$.
Hence $\Phi$ is a ground state of of $H(\lambda)$ and
$\langle \Phi|H(\lambda)|\Phi\rangle$ is the minimum
of the functions
$\lambda\mapsto\langle \phi_i|H(\lambda)|\phi_i\rangle,\;i=0,\ldots,k$.
This is impossible unless these functions coincide for all $\lambda\in{\Bbb R}$.
It follows that ${\Bbb J}_1\in\bigcap_{i=0}^k {\cal C}_{\phi_i}\equiv K$.
Hence $F$ is contained in the standard face $K$. This holds also in the case of $F$ consisting of a
single extremal point. \\
If $F$ is not equal to $K$ it must be part of its boundary since $F$ is a face. If ${\Bbb J}$ is an
interior point of $F$, as before, it must lie at the boundary of $K$ and, by lemma \ref{L7}, must
be contained in a smaller standard face $K'= \bigcap_{i=0}^\ell {\cal C}_{\phi_i},\; \ell>k$.
This is a contradiction since the ground states of $H({\Bbb J})$ are $\phi_0,\ldots,\phi_k$, see
above.\\
Thus $F=K$, i.~e.~every face of ${\cal C}_\Phi$ is a standard face and the proof of theorem \ref{T4}
is complete.
\end{enumerate} \hspace*{\fill}\rule{3mm}{3mm}  \\
It will often be convenient to represent the cone ${\cal C}_\Phi$ by its intersection
${\cal K}_\Phi={\cal C}_\Phi\cap P$ with a suitable hyperplane $P$, see e.~g.~the example in section
\ref{sec4.2}. It follows that the faces of ${\cal C}_\Phi$ are in $1:1$ correspondence with
the faces of the closed convex set ${\cal K}_ \Phi$,
except for the vertex ${\Bbb J}=0$ of the cone ${\cal C}_\Phi$.
The boundary of ${\cal K}_\Phi$ is partly flat, consisting of faces with dimension $d\ge 1$,
partly curved, consisting of extremal points ${\Bbb J}$ of ${\cal K}_\Phi$ not contained in larger faces
except ${\cal K}_\Phi$ itself. For these extremal points ${\Bbb J}$ the ground state of
$H({\Bbb J})$ is two-fold degenerate: $\{{\Bbb J}\}=\bigcap_{i=0}^1 {\cal C}_{\phi_i}$.
It is already maximally degenerate in the sense that any smaller face $\bigcap_{i=0}^2 {\cal C}_{\phi_i}$
will be empty.\\
Finally we note that throughout this section we never used the special structure of $\Phi$ as a product
of AF dimer ground states. Hence our analysis will also apply to the cones ${\cal C}_\phi$ defined
by other ground states $\phi$, but probably the
dimerized ground state cone ${\cal C}_\Phi$ will show the richest structure compared with other cones
${\cal C}_\phi$.

\section{Summary}
\label{sec7}
In this article the classes ${\cal C}_\Phi^s$ of DGS systems have been
investigated and completely characterized in the classical case $s=\infty$.
In the quantum case we calculated the linear space ${\cal J}_\Phi$
generated by ${\cal C}_\Phi^s$, but ${\cal C}_\Phi^s$ itself could only be explicitely determined
for small $N$ and $s$. In the general case, a ``lower bound" ${\cal C}'\subset {\cal C}_\Phi^s$
was constructed as the convex hull of three (resp.~four) special subsets of
${\cal C}_\Phi^s$ for arbitrary $s$ (resp.~$s=1/2$).
The facial structure of the convex cone ${\cal C}_\Phi^s$ turned out to be
anti-isomorphic to the lattice of eigenspaces of ground states of
$H({\Bbb J}),\; {\Bbb J}\in {\cal C}_\Phi^s$. \\
These results could be used for different purposes: They could
help to identify concrete spin systems as systems possessing
dimerized ground states or to better understand known examples of
DGS systems. A considerable improvement of the given lower bound
${\cal C}'\subset {\cal C}_\Phi^s$ seems to be difficult. On the
other hand, this article contains elements of a ``geometry of
multi-dimensional level crossing" which could also be interesting
in the broader context of quantum phase transitions. It seems
worth while and possible to extend the results on the geometrical
structure of ${\cal C}_\Phi^s$. For example, it remains an open
problem to prove or disprove the conjecture
${\cal C}_\Phi^s\subset {\cal C}_\Phi^{s'}$ if $s>s'$.

\section*{Acknowledgement}
I thank  K.~B\"arwinkel, M.~Luban, J.~Richter, and J.~Schnack for
stimulating and helpful discussions, J.~Richter for pointing out
some relevant literature, K.~B\"arwinkel for critically reading
the manuscript and M.~Kadiroglu for technical assistance in
preparing figure \ref{fig05}.

\end{document}